# Are Alien Civilizations Technologically Advanced?

The answer may depend on exo-planet politics

_____

By Abraham Loeb on December 26, 2017

As we discover numerous habitable planets around other stars in the Milky Way galaxy, including the nearest star, *Proxima Centauri*, one cannot help but wonder why have we not detected evidence for an advanced alien civilization as of yet. As the famous physicist Enrico Fermi asked: "where is everybody?" Even though the first interstellar object to be discovered in the solar system, 'Oumuamua, had an unusually elongated shape as might be expected from an alien probe, it does not appear to maneuver and is radio quiet below the level of a single cell phone.

True, a signal from an alien civilization might be subtle or sophisticated, but the disappointing silence of the sky may also indicate that old civilizations do not use technologies that would make them visible to our telescopes.

Based on our own experience, we expect old civilizations to be scientifically savvy and hence technologically advanced. But it is also possible that a primitive lifestyle rather than scientific prosperity has dominated the political landscape on other planets, leading to old civilizations that are technologically primitive.

Could exo-planet politics explain Fermi's paradox?

Human history allows us to imagine the possibility that our civilization could have remained dominated by the mindset of the middle ages under a different political scenario. Although such a scenario is imaginable over the timescale of thousands of years, its likelihood to prevail over millions or billions of years is unclear. Perhaps we were very lucky to mature as a technological civilization (in the spirit of the novel "Origin" by Dan Brown). Environmental or political disasters could have easily reset our evolutionary clock.

Or perhaps the ultimate lifetime of our advanced civilization will turn out to be shorter than it would have been if we remained technologically primitive because of the long-term risks that our technology poses to our future in terms of climate change or non-conventional (nuclear, biological or chemical) wars. In this case, the surfaces of other planets will show either relics of advanced civilizations that destroyed themselves by self-inflicted catastrophes or living civilizations that are technologically primitive.

We could search for the remnants of technological civilizations from afar. But if we detect nothing through our telescopes, the only way to find out whether long-lived civilizations are technologically primitive is to visit their planets. Astro-sociology could become a particularly exciting frontier of exploration as we venture into space.

Traditional astronomers would argue that it is much less expensive to remotely observe distant planets than to launch a probe that will visit them. But this argument misses the point that remote observing can only detect civilizations that transmit electromagnetic signals; change the planet's atmosphere (through industrial pollution or climate change); or leave artifacts on the planet's surface (in the form of photovoltaic cells, industrial infrastructure, artificial heating or artificial illumination).

If the aliens do not modify their natural habitat or transmit artificial signals, we will be forced to visit their home planets in order to uncover their existence. Non-technological civilizations might mesh seamlessly with their natural environment for a variety of reasons. At a minimum, camouflage is a natural survival tactic, and so primitive civilizations might prefer to appear indistinguishable from rudimentary forms of life, such as vegetation. But one could also imagine a civilization so intelligent that it deliberately keeps a low-key technological profile to sustain its biosphere, maintaining a lifestyle reminiscent of Henry Thoreau on Walden Pond. The only way to find these extraterrestrials would be to send probes that will visit their planets and report back.

The first significantly funded project to visit another planetary system, *Breakthrough Starshot*, was inaugurated in 2016. The *Starshot* concept was defined with the goal of reaching the nearest stars within a couple of decades. Since even *Proxima Centauri* is 4.24 light years away, this necessitates a technology that is capable of launching a spacecraft to at least a fifth of the speed of light. The only suitable concept involves a lightweight sail (to which the payload is attached) pushed by a powerful beam of light. But the downside of reaching such a high speed in this concept is that braking near the target planet is not feasible without a similar beamer there.

Visiting the surface of another planet therefore requires slower speeds and longer travel times. For example, conventional rockets would bring us to the nearest stars within hundreds of thousands of years. This might still be appealing from an academic perspective, since this timescale is tens of thousands times shorter than the age of the Universe. Over the billions of years available to our technological civilization to explore the Milky Way galaxy, we could compile a sociological census of billions of exo-planets. And even if we find mostly faith-based alien cultures instead of advanced infrastructure that would accelerate our own technological development, it would be fascinating to explore the diversity of Galactic interpretations for the concept of God.

## ABOUT THE AUTHOR

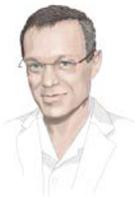

**Abraham Loeb**

Abraham Loeb is chair of the astronomy department at Harvard University, founding director of Harvard's Black Hole Initiative and director of the Institute for Theory and Computation at the Harvard-Smithsonian Center for Astrophysics. He also chairs the advisory board for the Breakthrough Starshot project.

Credit: Nick Higgins